# Correlation of the role of Li-doping in control of O-vacancies and Li-interstitial formations in NiO with electrochemical properties

Poonam Singh[a], P. Maneesha[a], Manju Kumari[a], Abdelkrim Mekki[b,c], Khalil Harrabi[b,d], Somaditya Sen[a*]

[a]Department of Physics, Indian Institute of Technology Indore, Indore 453552, India

[b]Department of Physics, King Fahd University of Petroleum & Minerals Dhahran, 31261, Saudi Arabia

[c]Center for Advanced Material, King Fahd University of Petroleum & Minerals, Dhahran 31261, Saudi Arabia

[d]Interdisciplinary Research Center (RC) for Intelligent Secure Systems, King Fahd University of Petroleum & Minerals, Dhahran 31261, Saudi Arabia

[*]Corresponding author: sens@iiti.ac.in

## Abstract

Aliovalent doping in an oxide material introduces modifications in the valence state of the host cation and often leads to tailoring the oxygen content in the lattice. Moreover, if the dopant cation is larger than the host cation, the lattice strain and disorder may be affected. Such changes are expected to modify the electronic clouds and lead to different ligand fields, which in turn should modify the bond lengths, and therefore phonons, electronic properties, transport properties, and charge storage properties. To understand such correlations an example is being investigated in this study by doping a larger $Li^+$ ion in a NiO lattice. The effect on structure, phonons, electronic properties, and charge storage properties are investigated and correlated in a first-of-its-kind report. The charge storage properties are observed to improve with $Li^+$ doping until 3% substitution and thereafter decrease due to the generation of $Li^+$ interstitial in a 6% incorporated sample. The connection of oxygen vacancies and $Ni^{3+}$ formation with $Li^+$ incorporation is the backbone of this report.

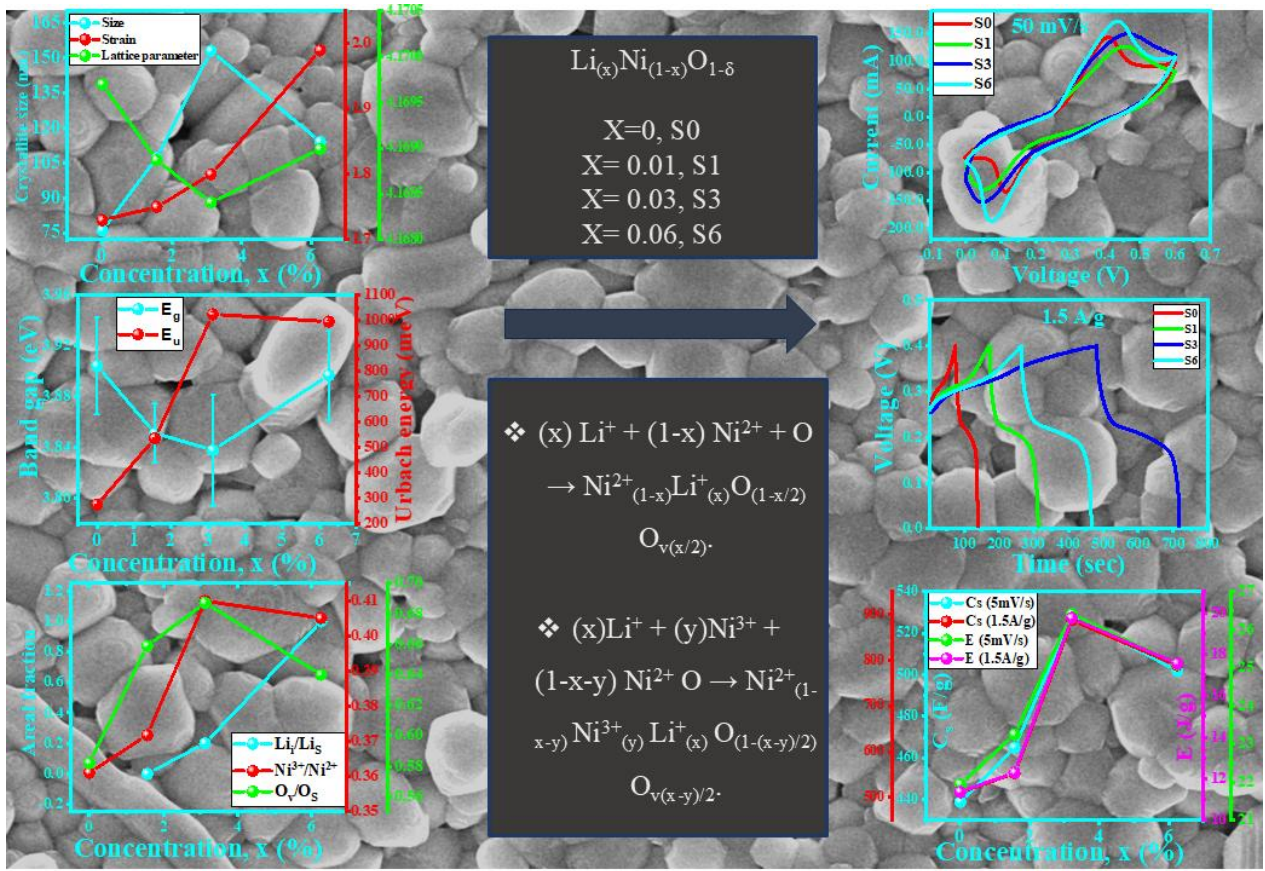

**Keywords**

Oxygen vacancy, Li interstitials, Strain and Disorder, Structure Correlated Physical properties, Conduction properties, Electrochemical Properties

## 1. Introduction

The modern-day has become critical on energy requirements and there is a serious energy crisis that requires new energy harvesting from renewable sources and efficient energy storage devices like batteries, fuel cells, and supercapacitors [1, 2]. The most effective energy storage device among all existing systems is the electrochemical supercapacitor, which has strong cycle stability, high power as well as high energy density [3, 4]. Nickel oxide (NiO) is a well-known capacitive electrode material known for its thermal stability, high chemical stability, low price, natural abundance, and environment friendliness [5]. It has a high theoretical capacity [6]. However, in practice, it shows the low value of specific capacitance as the un-doped NiO is an insulator (~ $10^{13}$ Ω cm) [7-9]. To improve the transport properties a lot of chemical modifications have been tried until now including substitution by transition metal and rare earth elements [10-13]. Different morphologies have also been tried on different types of doping [14-20]. Improvement of pseudo-capacitance and energy storage properties have also been revealed by several studies [14-22]. Although the effect of a lower valence state dopant like $Li^+$ has been studied sparsely in different types of reports [22-33, 35-39], one of the important aspects that have not been discussed well is a probable mechanism and correlation of how the lower charge on the cation can modify the structural and electronic properties of the material, thereby changing the transport properties and electrochemical properties [23, 24, 28-30]. Also, one of the important aspects that is observed in literature is the high percentage of doping concentration (sometimes ~30%) which seems to be challenging in practice [2, 11, 14, 23, 29-33]. This is because the size of a $Li^+$ ion (0.9 Å) is much larger than the host $Ni^{2+}$ ion (0.83 Å) [34], which should introduce a huge lattice strain and thereby hamper the crystallinity and phase of the material. A lesser valent and larger size ion like $Li^+$ will have a solubility limit in a robust structure like NiO [35]. Hence, reports with such a high amount (>3%) of substitution remain a question mark. There are reports with lower percentages (<3%) [36-39] which seem more probable and realistic. However, even these reports fail to explain the changes in the lattice due to $Li^+$ substitution ($Li_s$) and correlate the changes to the properties that have not been discussed in the literature. Hence, the consequences of such doping and the limits need to be investigated and justified. There can be several consequences of such a lower valent $Li^+$ incorporation in place of $Ni^{2+}$ in the NiO lattice to preserve the charge neutrality, like a transformation of $Ni^{2+}$ to $Ni^{3+}$ [20], the creation of oxygen vacancies ($O_v$), and the creation of $Li^+$ interstitial ($Li_i$). Such changes can definitely modify the ligand field and hence the bonds, thereby changing the electronic properties and hence transport [22] and charge storage properties. This study is a journey of understanding the limits of $Li^+$ incorporation and correlating the structure, vibrational properties, electronic properties, and charge storage properties.

## 2. Experimental procedure

$Ni_{(1-x)}Li_{(x)}O$ nanocrystalline powders were synthesized using nickel (II) nitrate hexahydrate ($Ni(NO_3)_2.6H_2O$, 99%, Alfa Aesar), and lithium(I) nitrate ($LiNO_3.9H_2O$, 99%, Alfa Aesar) as precursors. Calculated amount of the above chemicals was taken in separate beakers and dissolved in doubly ionized water (DIW). After obtaining homogeneous solutions, the lithium solution was added to the nickel solution. The mixture was magnetically stirred for a few hours to insure the homogenous mixing of the two precursors. In another beaker a calculated amount of citric acid and ethylene glycol were mixed thoroughly in DIW. The clear citric acid and ethylene glycol solution was added to the precursor solution to act as monomer and thereafter polymer formation upon uniformly heating at 80 °C on a hot plate required for gel-formation. This polymer network will later on act as a fuel during burning of the gel at the same temperature. The obtained powders were then dried and decarbonized and denitrification at 450 °C. Further heated at 600 °C to finally obtain fluffy black powders containing nanoparticles of $Ni_{(1-x)}Li_{(x)}O$. The samples were named S0 (for x = 0), S1 (x=0.016), S3 (x=0.031), and S6 (x=0.062) [Fig. 1].

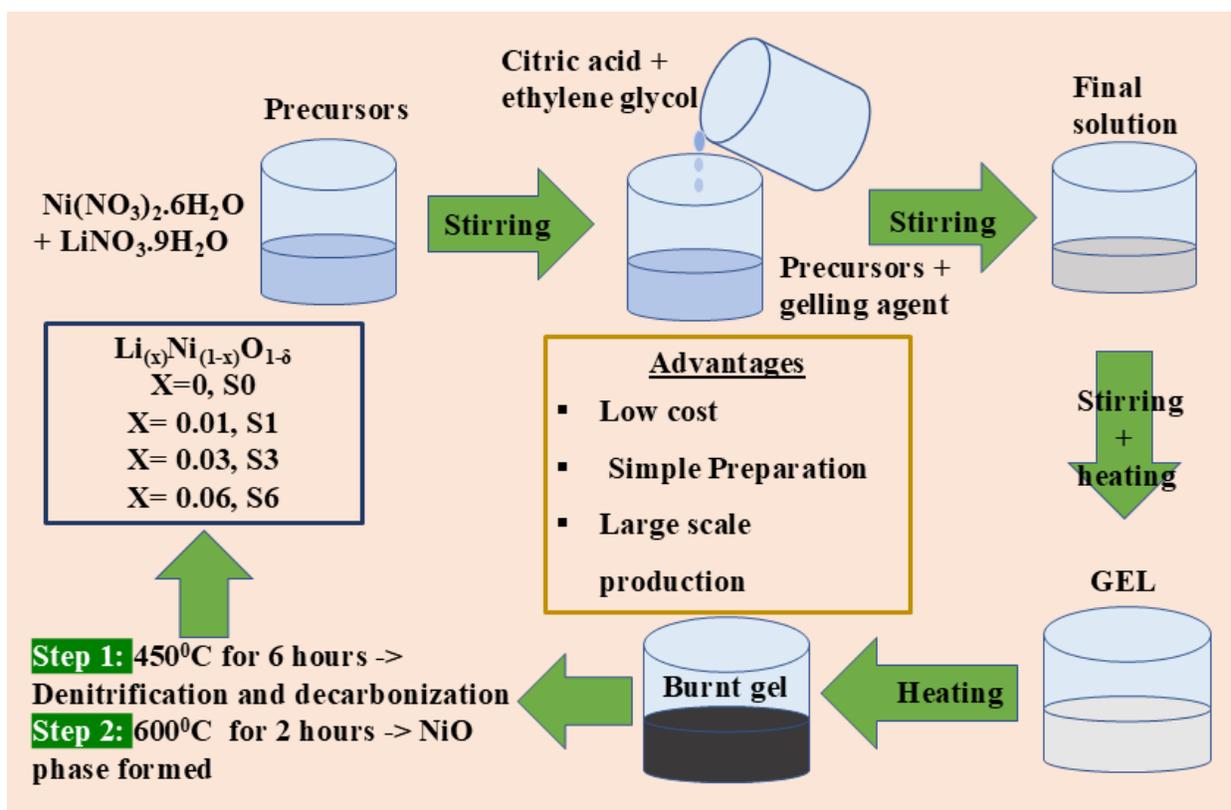

*Figure 1: Graphical representation for the synthesis of Li-doped Ni via sol-gel synthesis route.*

The surface morphology of the samples was studied using a Carl Zeiss FESEM/EDS-Supra55 microscope. The structural phase of these samples was confirmed using a Bruker D2 Phaser diffractometer equipped with Cu-Kα radiation ($\lambda = 1.54$ Å). The vibrational properties, i.e., the phonons were studied using a LabRAM HR Evolution (Horiba) Raman spectrometer. A He–

Ne laser of wavelength *632.8* nm was employed for excitation. The electronic band gap was obtained from the UV-Vis diffuse reflectance spectra (DRS) using a Perkin Elmer Lambda *35* UV-Visible spectrophotometer. A Teflon (polytetrafluoroethylene, or PTFE) sample was used as a reference material for perfect reflection. For X-ray photoelectron spectroscopy (XPS) measurements using Thermo-Scientific Escalab *250* Xi, monochromatic Al $K_α$ X-rays ($hv$ = *1486.6 eV, λ = 8.3 Å*) operating at *150W* under ultra-high vacuum (~$10^{-9}$ mbar) were employed. A survey spectrum was initially obtained for each sample in the binding energy range of *0~1400* eV. Thereafter, high-resolution Ni2p (*840~890* eV), Li1s (*46~62* eV), and O1s (*520~540* eV) spectra were obtained. The spectra were analyzed with the help of XPSPEAK4.1 software.

Cyclic voltammetry (CV) measurements were carried out using an electrochemical workstation (make: Kanopy). An Ag/AgCl (3M KCl) reference electrode was used with platinum wire as the counter electrode. Ni foams (NFs) (Sigma-Aldrich: thickness of 2 mm, bulk density of 0.45 g cm$^{-1}$, porosity of 95%, 20 pores/cm, and purity of 99.5%) were used as the working electrodes onto which the active material was deposited by drop-casting a dispersed solution containing the materials. The NFs were cut into strips of dimension 2cm × 1cm. An optimization was performed on the effect of the etching of NFs. It was observed that a short time etching (≤1min) with a low HCl concentration (≤0.1M) is best suited for removing contamination and increasing the roughness thereby improving the adhesive force between substrate and active material. For a more concentrated HCl solution (> 0.1 M) and longer etching time (> 1 min) an oxide layer seemed to form on the NFs. This can lead to additional NiO layers of different properties. Hence such conditions were avoided by sonicating the NFs in 0.1 M HCl for only 1 min for this study. These etched NFs pieces were thereafter cleaned with ethanol. The Li-doped NiO samples were deposited on these cleaned NFs substrates for electrochemical performances. Different concentrations of the active material (Li-doped NiO) were prepared by mixing 2mg, 3mg, 4mg, 5mg, and 6mg of the active material with acetylene black, and Nafion in a weight ratio of 80:10:10 in 500 μL of ethanol. The combination was ultrasonically mixed for one hour to obtain dispersed solutions with different concentrations. The dispersed solutions were drop-casted on the NFs with care such that the entire NF gets evenly coated. These coated foams were left to dry in an oven overnight at 80°C [Fig. 2]. After drying overnight at 80°C, the coated NFs were ready for CV and Galvanostatic Charge-Discharge (GCD) studies. The stability of the coats was verified before performing CV and GCD. The electrochemical properties (CV and GCD) were studied using a three-electrode assembly cell with a 6M KOH electrolyte within a positive potential window.

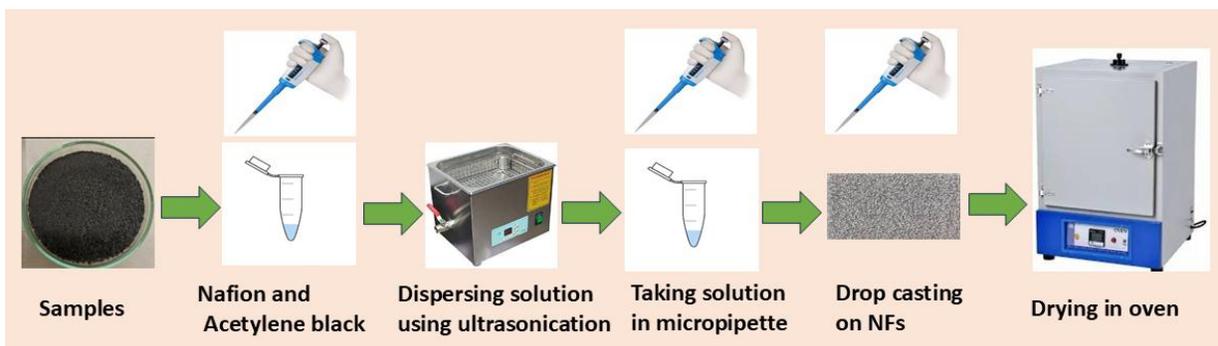

*Figure 2: Graphical representation for the working electrode preparation via drop casting method on NFs.*

## 3. Results and Discussion
### 3.1. FESEM and EDS

FESEM studies were performed to investigate the surface morphology of the samples [Fig. 3]. The effect of $Li^+$ on the particle size and morphology is important to understand the charge storage properties. FESEM images of the doped samples reveal an increase in size as compared to the pure NiO nanoparticles. However, the morphology remains more or less the same, with agglomerated spherical nanoparticles forming the building block of the material. The agglomerations seem to be a collection of larger nanoparticles with a coat of numerous extremely small nanoparticles/nanodots. This gives the morphology a granular and porous nature [24, 26-27].

The $Li^+$ content is low enough to be detected. Hence, it was extremely difficult to estimate the $Li^+$ content from the EDS spectra [Table 1]. Hence, there were no particular changes in the EDS results for the Li-doped samples. EDS contains only Ni and O in a ratio of 2:1. Since EDS provides data in weight %, Ni (58.69 u) contributes more to the total weight than O (∼ 16 u).

*Table 1: Obtained Weight % of elements calculated from EDS measurement*

| Element | S0 | S1 | S3 | S6 |
| --- | --- | --- | --- | --- |
|  |  |  |  |  |
| Li | 0.0 | 0.0 | 0.0 | 0.0 |
| O | 33.4 | 35 | 34.8 | 35.4 |
| Ni | 66.6 | 65.0 | 65.2 | 64.6 |

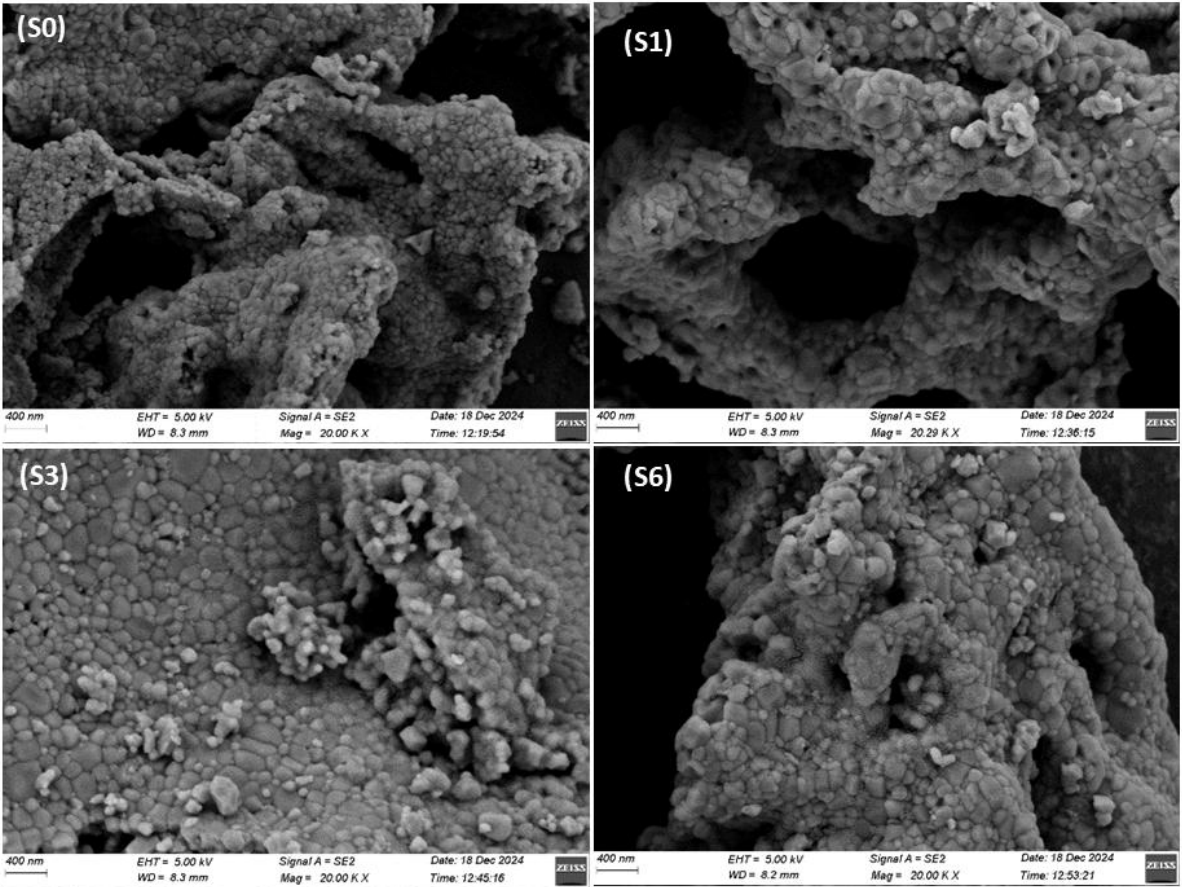

*Figure 3: FESEM images of Li-doped NiO samples revealing agglomerated nanoparticles with 20k magnification.*

### 3.1. X-ray diffraction (XRD) Analysis

The XRD patterns of the Li-doped NiO nanocrystalline powders revealed diffraction peaks corresponding to $Fm\bar{3}m$ structure (COD #1010093) [Fig. 4 (a)] [40]. Rietveld refinement [Fig. 4 (b), Supplementary Fig. S1] revealed a decrease in lattice parameters for Li-doped samples compared to pure NiO. With the increase in Li-incorporation, the lattice parameters decreased from 4.17 Å (S0) to 4.169 Å (S1) and 4.168 Å (S3) [35], and thereafter slightly increased to 4.17 Å (S6) [Figure 4 (c)] [35]. The unit cell volume decreased accordingly from 72.512 Å$^3$ (S0) to 72.459 Å$^3$ (S1) and 72.407 Å$^3$ (S3), and thereafter slightly increased to 72.511 Å$^3$ (S6) [24, 30]. In NiO, the oxidation state of Ni is supposed to be in the Ni$^{2+}$ state with coordination six. The ionic radii of Ni$^{2+}$(VI) ion are 0.83 Å [34]. With the incorporation of Li$^+$, one can envisage several scenarios involving different vacancies, interstitials, valence states, etc. However, the most important and natural consequence of the presence of a lower valence Li$^+$ in place of a higher valence state Ni$^{2+}$ is expected to create O$_v$. Hence, this may lead to a lattice containing Ni$^{2+}$, Li$^+$, O$^{2-}$ and O$_v$ according to the following possibility [26, 32]:

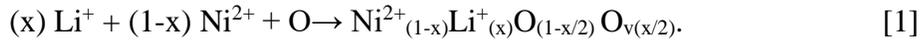

$$(x)\,Li^+ + (1-x)\,Ni^{2+} + O \rightarrow Ni^{2+}_{(1-x)}Li^+_{(x)}O_{(1-x/2)}\,O_{v(x/2)}. \qquad [1]$$

However, the presence of Li$^+$ can also induce a transformation of a Ni$^{2+}$ ion to a Ni$^{3+}$ valence state, thereby reducing the total O$_v$ content [9, 26, 37-38]:

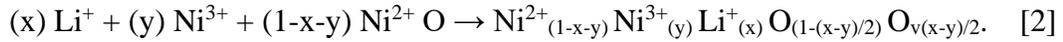

$$(x)\,Li^+ + (y)\,Ni^{3+} + (1-x-y)\,Ni^{2+}\,O \rightarrow Ni^{2+}_{(1-x-y)}Ni^{3+}_{(y)}Li^+_{(x)}\,O_{(1-(x-y)/2)}\,O_{v(x-y)/2}. \qquad [2]$$

Such possibilities can only be verified by studying the valence states of the constituent elements using probes like XPS. Note that the Ni$^{3+}$(VI) ion is much smaller ~ 0.74 Å (high spin, HS) or ~ 0.7 Å (low spin, LS) than the Ni$^{2+}$ ion, while the Li$^+$(VI) ion are much larger (0.9 Å) than both oxidation states of Ni ions [34]. A simple substitution of Ni$^{2+}$ by Li$^+$ should expand the lattice. This is contradictory to the unit cell volume contraction in these samples S1 and S3 [39]. Hence, the volume contraction in S1 and S3 may be a consequence of O$_v$ and Ni$^{3+}$ formation [20]. In case of S6 the volume expansion can be a consequence of some Li$^+$ ions occupying interstitial positions which requires extra space in the lattice [20]. To verify and substantiate these claims XPS analysis becomes an important study in this work. Such inclusion of O$_v$, Li$_i$, Ni$^{3+}$ ions are not regular events for the lattice, and may result in the deformation or irregularity of the lattice, thereby increasing the lattice strain and disorder [24]. The microstructural strain ($\varepsilon$) and crystallite size ($D$) were estimated using the Williamson–Hall (WH) equation [42]: [$\beta.Cos\theta = \varepsilon.4Sin\theta + K\lambda/D$; a "y =

$mx+c$" format where y = $\beta \cdot \cos\theta$, x = $\sin\theta$, m = $\varepsilon$, and c = $K\lambda/D$], where, $\beta$ is full width at half maxima (FWHM), $\theta$ is angle of incidence of X-ray, $\lambda$ is wavelength of X-ray (*1.5406 Å*), and K (~ *0.94*) is a constant dependent on morphology. The $\varepsilon$ was observed to increase with substitution from 0.00173 in S0 to 0.00175 in S1, 0.0018 in S3 and 0.00199 for S6 [24] [Fig 4 (d)]. Note that this continuous increase of strain indicates the effect of the larger Li$^+$ ion in the lattice is associated with a consequential introduction of Ni$^{3+}$, O$_v$, and Li$_i$ [24]. However, D increased from S0 (76 nm) to S1 (107 nm), and S3 (153 nm) [24], and thereafter decreased for the S6 sample (114 nm). Hence, the crystallites are large enough for the chemically modified samples to be called nanoparticles. However, the crystallinity seems to increase for S1 and S3 samples with the increase of size of the crystallites which decrease for the S6 sample, may be due to an interstitial nature of Li, which increased the strain further [20, 24, 35].

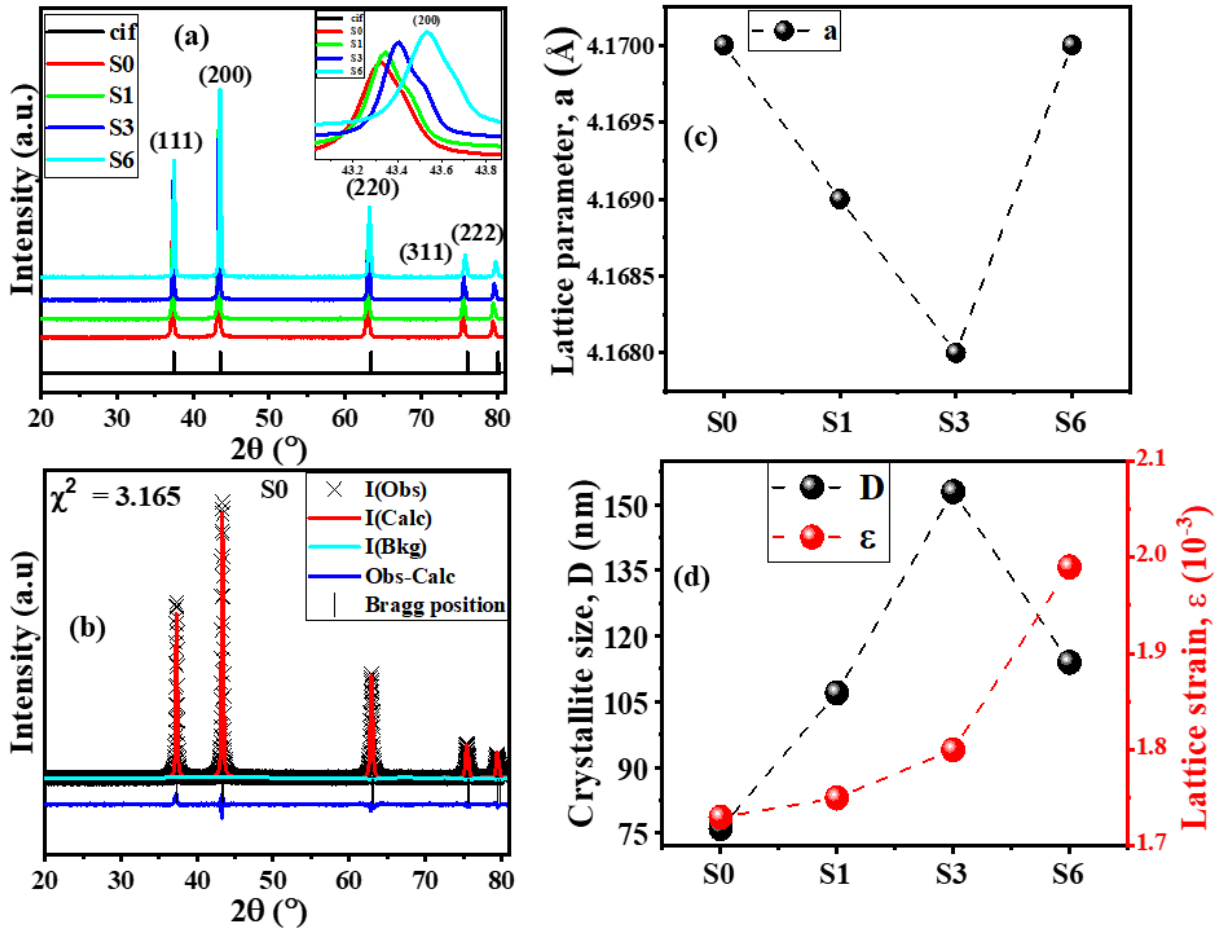

*Figure 4 :(a) XRD pattern of all samples, (b) Rietveld-refined pattern for pure sample, (c) Variation of lattice parameter with substitution, (d) variation of crystallite size and strain with substitution calculated from Williamson–Hall (WH) formula.*

## 3.2. X-ray Photoelectron Spectroscopy (XPS) Analysis

XPS technique is employed to further confirm the chemical composition and valence states of the synthesized samples. The Ni-2p, O-1s, and Li-1s spectra at room temperature are analyzed to understand these properties. For the pure NiO sample, the Ni-2p spectra revealed seven convoluted peaks, belonging to $Ni^{2+}$-$2p_{3/2}$ (853.15 eV), $Ni^{3+}$-$2p_{3/2}$ (854.94 eV), $Ni^{2+}$-$2p_{1/2}$ (870.45 eV), $Ni^{3+}$-$2p_{1/2}$ (872.24 eV), with three satellite peaks (at 860.29 eV, 865.01 eV, and 878.42 eV) [20, 43] [Fig. 5 (a, d), Supplementary Fig. S2]. Therefore, the spin-orbit splitting (SOS) energy between the $Ni^{2+}$-$2p_{3/2}$ and $Ni^{2+}$-$2p_{1/2}$ states is observed to be ~17.3 eV. An exactly similar SOS ~17.3 eV was found between the $Ni^{3+}$-$2p_{3/2}$ and $Ni^{3+}$-$2p_{1/2}$ states. This binding energy difference of ~17.3 eV is close to the standard value (17.49 eV) [9, 44-45]. An estimate of the relative quantities of $Ni^{2+}$ and $Ni^{3+}$ content was calculated from the peak areas of the two contributions in the Ni-2p spectra [Fig 5 (f)]. With an increase of Li concentration an increase in $Ni^{3+}/Ni^{2+}$ peak area fraction is observed. The $Ni^{3+}/Ni^{2+}$ peak area fraction increases from 0.361 (S0), 0.372 (S1) to 0.41 (S3) [20], and finally nominally reduces to 0.405 (S6). This hints at $Ni^{2+} \rightarrow Ni^{3+}$ conversion with Li doping until S3 with a slight reduction in conversion for S6. A slight shift towards lower binding energy is observed in both Ni-$2p_{3/2}$ and Ni-$2p_{1/2}$ peaks with Li doping, which indicates that the electron density around the cations is reduced [43].

For the pure NiO sample, the O-1s spectra revealed three convoluted peaks, belonging to $Ni^{2+}$-O (528.87 eV, denoted as $O_I$), $Ni^{3+}$-O (530.62 eV, denoted as $O_{II}$), and water adsorbed (532.32 eV, denoted as $O_{III}$) [11, 43] [Fig 5 (b, e), Supplementary Fig. S3]. The $Ni^{3+}/Ni^{2+}$ ratio was also estimated from the peak areas of the $Ni^{2+}$-O and $Ni^{3+}$-O peaks [Fig 5 (f)]. The $Ni^{3+}/Ni^{2+}$ ratio from this analysis also increases with Li doping, in agreement with the results of Ni-2p spectra.

For the maximum Li-doped samples, the Li-1s spectra revealed two convoluted peaks, belonging to $Li_i$ (~53.8 eV), and $Li_s$ (~55 eV) [9] [Fig 5 (c), Supplementary Fig. S4]. However, these features are noisy for all the samples, especially for the lower percentages. $Li_i$ typically shows a lower binding energy compared to $Li_s$, reflecting a more "metallic" or less bound state [20]. Note that the $Li_i$ contribution is not observed for S1, i.e., for the least lithium-doped sample. This peak emerges and becomes stronger with increasing Li content. The ratio of $Li_i/Li_s$ increases from 0 (S1) to 0.253 (S3), and finally to 1.02 (S6). It is observed for S6 this ratio is considerably high and almost 50% of the Li ions are $Li_i$. For the other Li-doped samples the number is not that significant [Fig 5 (f)].

The lattice Li is expected to introduce oxygen deficits in its vicinity (as followed by the above equations [1] and [2]). Hence from the Ni-2p and Li-1s analysis, it seems that simultaneous increase of $Li_s$ and $Ni^{3+}$ happens, trying to balance a drastic increase of $O_v$. However, the amount of $Li_s$ seems to dominate over $Ni^{3+}$ thereby enhancing the $O_v$ content until S3. On the other hand, the simultaneous creation of $Li_i$ in S3 and S6 reduces the amount of $O_v$ further, especially for the S6 sample where $O_v$ content is reduced due to excessive $Li_i$ [Fig 5 (f)]. The indications from the XRD results about a possible conversion of $Ni^{2+}$ to $Ni^{3+}$ and thereby its contribution to $O_v$ are better understood from the XPS studies, and are mutually in corroboration with each other.

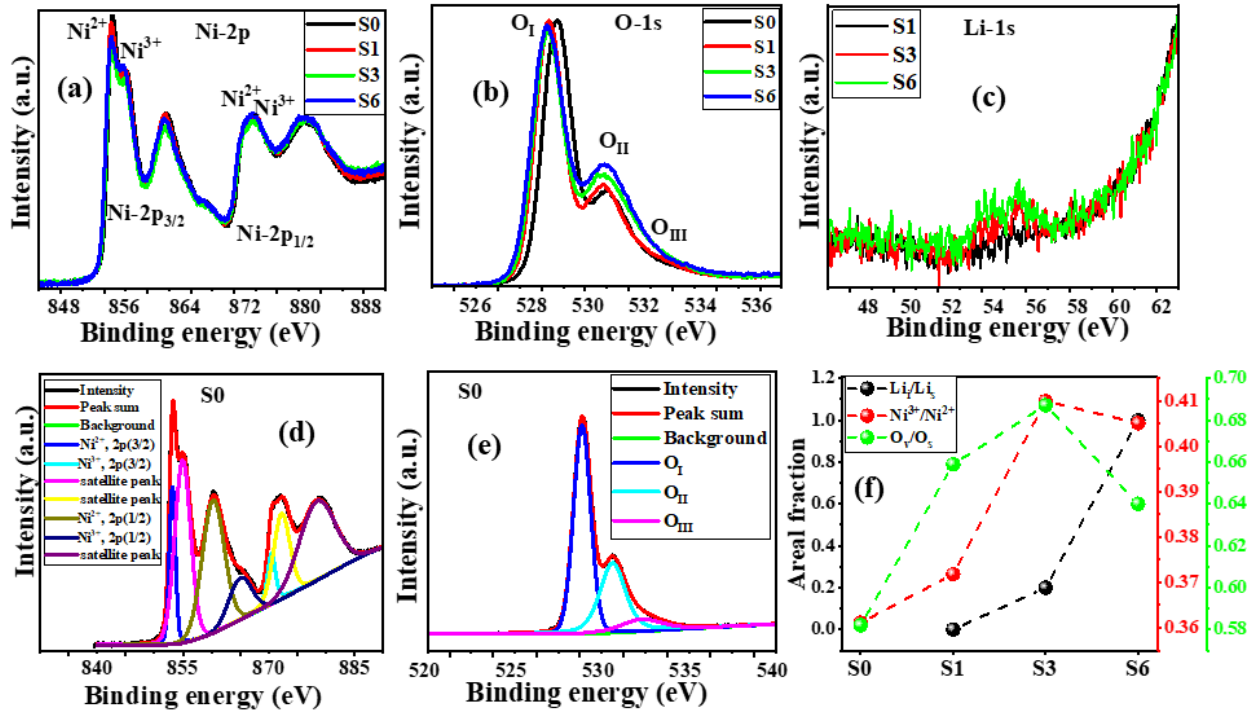

*Figure 5: (a) XPS spectra of Ni-2p peaks for all samples, (b) XPS spectra of O-1s peaks for all samples, (c) XPS spectra of Li-1s peaks for all samples, 3(d) and 3(e) fitted XPS spectra of S0 sample for Ni and O respectively, (f) Variation of the $Ni^{3+}/Ni^{2+}$ peak area fraction with substitution.*

### 3.3. UV-Visible Diffuse Reflectance Spectra Analysis

The band gap, $E_g$, was estimated from the reflectance percentage ($R\%$) obtained from UV-Vis Diffuse Reflectance Spectra (UV-Vis DRS) in the wavelength regime 200-1100 nm. Thick non-transparent samples were used to reduce the transmission coefficient to zero, allowing only the possibilities of absorbance (*A%*) and reflectance (*R%*). The Kubelka-Munk function, *F(R)* [24, 45] was used to calculate the equivalent absorption coefficient *(α)*, $\alpha \cong F(R) = (1-R)^2/2R$, where *R (=R%/100)* is the diffuse reflectivity. Using the Tauc method [46], the optical band gap was computed using energy-dependent optical absorbance data. The Tauc method involves a relationship between *α*, energy *(E) = hv*, and $E_g$ in the form: $(\alpha h\nu)^{1/n} = A(h\nu - E_g)$; where *α* is the absorption coefficient, *v* is the photon's frequency, *h* is Planck's constant, *A* is a proportionality constant, and $E_g$ is the band gap. For a direct allowed bandgap, *n=½* [46]. Hence, a Tauc plot of $(\alpha h\nu)^2$ vs *hv* was plotted for all samples [Fig 6 (a)]. The optical $E_g$ was determined by protraction of the linear parts of this plot to the x-axis [$(\alpha h\nu)^2 = 0$]. The spectra of all samples exhibit one main absorption peak. However, multiple distinct pre-absorption features are observed for S0 and S1. The main absorption edge resulting in the energy gap, $E_g$ (band gap) is due to an O2p to Ni3d transition [14, 36, 42]. $E_g$ is observed to decrease with $Li^+$ doping from 3.86 eV in S0 to 3.85 eV in S1 and 3.84 eV in S3 [Fig 6 (b)] but increases to 3.89 eV in S6 [20, 24]. These changes are visually observed from Tauc plots. However, as observed from their values the changes are minimal and in the error range. Hence, not much significant claim can be made from these changes. However, with $Li^+$ doping an increase in the concentration of free charge carriers can be expected with the incorporation of lattice defects [14, 27, 29-30, reports showed that carrier concentration increased with Li-doping via Hall effect measurement] created due to a mismatch of the valence state and the ionic radii of the dopant and the host [Supplementary Fig. S5]. Hence, the visual differences in the primary absorption edge may be real.

The multiple distinct pre-absorption features in the visible spectra can be assigned to distinct absorption peaks originating from optical transitions between different energy levels of the d orbitals of $Ni^{2+}$ ions [24, 47]. A $Ni^{2+}$ ion in NiO has an electronic configuration of $3d^8$, in an octahedral field. In the crystal-field (CF) framework, the Hamiltonian for the transition-metal $3d^8$ ions can be expressed as [47-48], $H = H_{ee}(B,C) + H_{SO}(\xi) + H_{Trees}(\alpha) + H_{CF}(D_q)$. In the above expression, $H_{ee}$ (a function of the Racah parameters B and C) is associated with electron-electron repulsions which leads to the Russell-Saunders terms: the ground state $^3F$ (according to Hund rule), and the excited states $^3P$, $^1G$, $^1D$, and $^1S$. The term $H_{SO}$ represents the spin-orbit (SO) coupling, depending on the parameter $\xi$ (the spin-orbit coupling constant). On the other hand,

$H_{Trees}$ (function of the Trees parameter $\alpha$), is a correction term originating from a two-body orbit–orbit polarization interaction. The electron-electron interaction as in $H_{ee}$ is a modification of an electronic cloud and hence also associated with an orbit–orbit polarization. Hence $H_{ee}$ and $H_{SO}$ are often a result of a single change in the electronic cloud distribution. The $H_{CF}(D_q)$ component is dependent on $D_q$ (a CF strength parameter).

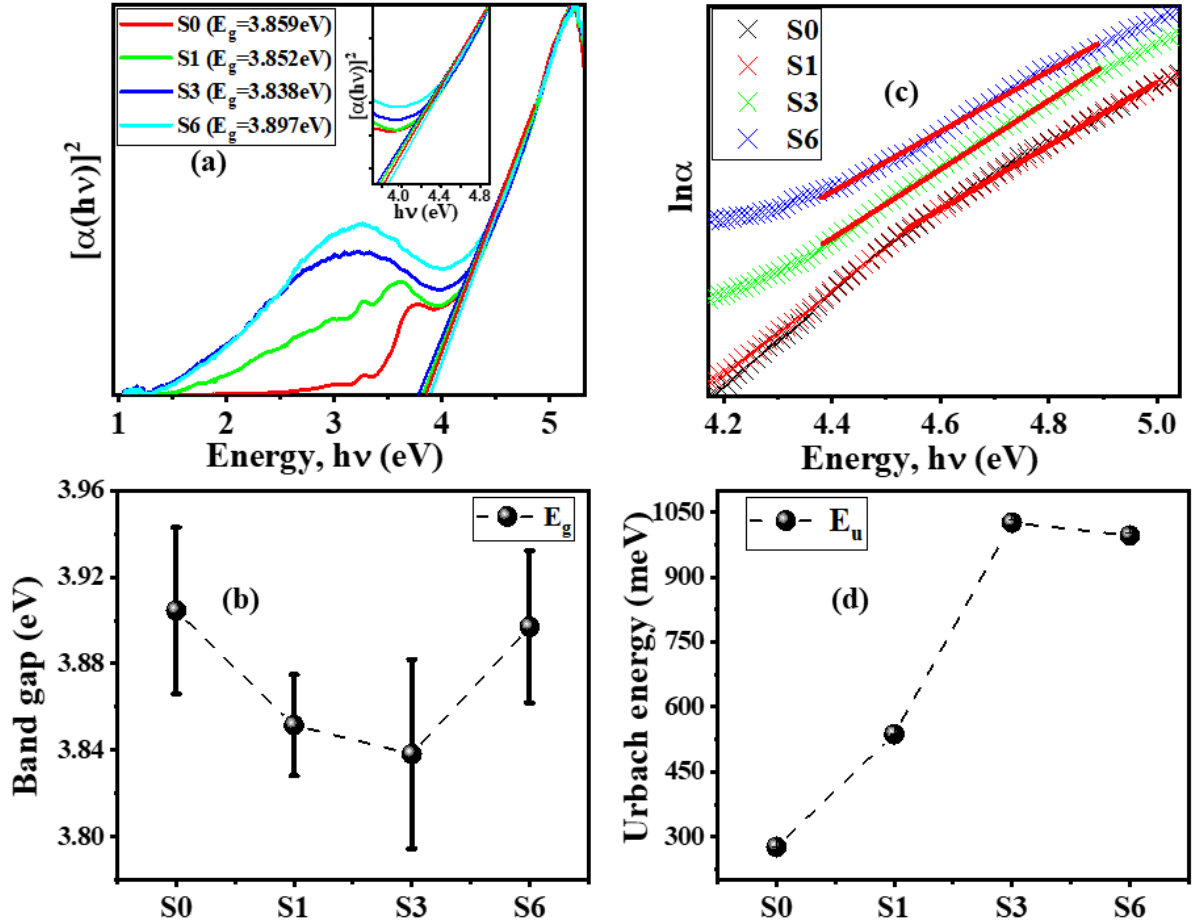

Figure 6: (a) Tauc plot of Li-NiO samples for the direct energy gap, (b) Variation of $E_g$ with substitution, (c) Linear fitting plots for $E_u$ calculation (d) Variation of $E_u$ with substitution

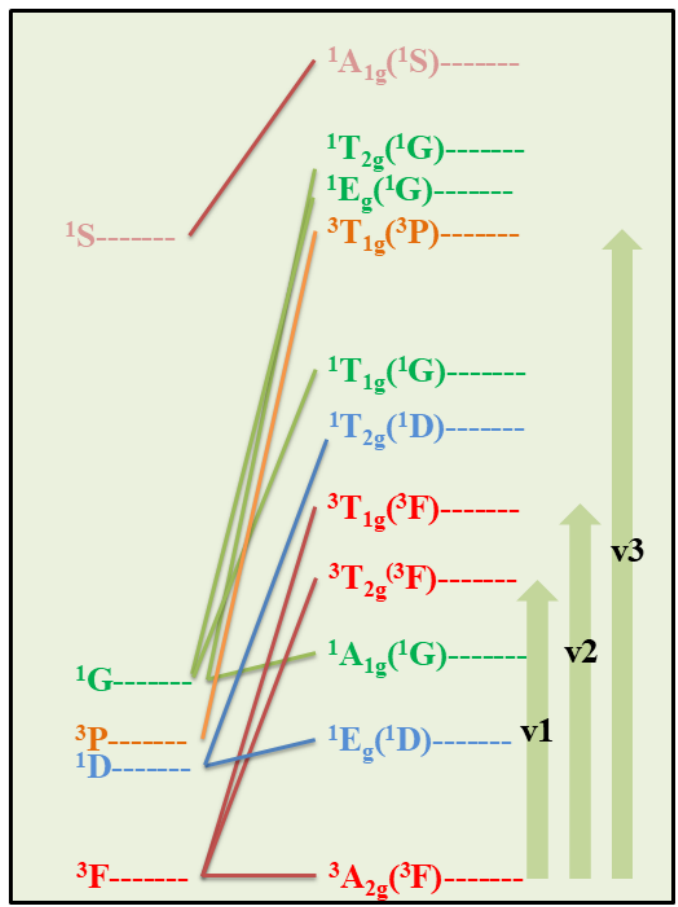

*Figure 7: Tanabe-Sugano diagram of a $d^8$ octahedral for $Ni^{2+}$ along with the allowed transitions [48].*

In an octahedral crystal field for a $d^8$-ion, Russell-Saunders energy levels split according to the strength of the octahedral ligand field. Note that the $^3F$ energy level splits into $^3T_{1g}$, $^3T_{2g}$, and $^3A_{2g}$ energy levels. Similarly, the $^1D$ energy level will split into $^1E_g$ and $^1T_{2g}$ energy levels while the $^3P$ energy level will give $^1T_{1g}$, and $^1S$ term will give $^1A_{1g}$. The above splitting can be represented by the Tanabe-Sugano diagram [Fig 7] for an extremely strong octahedral CF. Without any CF the energy levels are represented on the left-hand side of the diagram in the order $^3F<^1D<^3P<^1G<^1S$. In the strong octahedral CF, the energy levels are arranged in order $^3A_{2g}(^3F)<^1E_g(^1D)<^1A_{1g}(^1G)<^3T_{2g}(^3F)<^3T_{1g}(^3F)<^1T_{2g}(^1D)<^1T_{1g}(^1G)<^3T_{1g}(^3P)<^1E_g(^1G)<^1T_{2g}(^1G)<^1A_{1g}(^1S)$. Note that, all possible electronic transitions are not quantum-mechanically allowed. The allowed transitions are defined by two rules: the spin selection rule and the Laporte rule. According to the spin selection rule, transitions with unchanged total spin quantum number S (spin multiplicity) are allowed [36, 49]. On the other hand, the Laporte rule allows transitions with a change of parity ($g \Leftrightarrow u$) [49]. As a consequence of the above two rules only three electronic transitions (S=1⇔S=1) can be expected from $^3A_{2g}(^3F)$ (i.e. the ground state) to $^3T_{2g}(^3F)$, $^3T_{1g}(^3F)$, and $^3T_{1g}(^3P)$, which belong to energies below the bandgap. These three energy transitions are labeled as $v_1$, $v_2$, and $v_3$ respectively in Fig. 7 [48].

As discussed above, Li$^+$ incorporation in place of Ni$^{2+}$ introduces Ni$^{3+}$ along with O$_v$ [24]. Such modifications in the lattice may create disorder and defect states. Generally, disorder introduces modifications to the conduction band (CB) and valence band (VB) edges. Such modifications appear as band tails and are generally exponential in nature [27]: $\alpha = \alpha_0 \cdot exp(h\nu/E_u)$, where, $h\nu$ represents the photon energy. Here, the term $E_u$ represents an energy stored as a disorder in the lattice, known as Urbach energy. $E_u$ was calculated from the slope of $ln(\alpha)$ versus $h\nu$ plot [Fig. 6 (c)]. $E_u$ increases from S0 (0.276 eV) to S1 (0.537 eV) to S3 (1.026 eV) and thereafter slightly reduces for S6 (0.996 eV) [Fig. 6 (d)]. Hence, with Li$^+$ incorporation along with the increase of lattice strain, the lattice disorder also increases.

Hence, from the electronic properties study, it was observed that with Li$^+$ incorporation, the lattice gets strained and disordered hinting at modifications in the electronic clouds of the atoms and thereby changes in the bond lengths and probably bond angles. Such changes should affect lattice vibrations, transport properties, and charge storage properties.

### 3.4. Lattice vibration studies (Raman Spectroscopy)

The room temperature Raman spectra [Fig. 8 (a)] of the synthesized nanostructures reveal prominent broad features at 400 and 530 cm$^{-1}$. These correspond to the 1P(TO) and 1P(LO) vibrational modes of the NiO lattice, respectively. Some additional weak features are also observed at 1090 cm$^{-1}$ that can be assigned to a two-phonon 2P(2LO) vibrational mode of NiO [27].

Due to Li$^+$ doping in the NiO lattice, the notable changes can be linked with the changes in the effective mass, bond strength, vacancies, and electrostatic interaction between ions. The mass of the dopant Li (~ 6.94 u) is much lighter than Ni (58.69 u). As phonon frequency is directly related to the square root of the bond strength and inversely to the square root of the effective masses of the vibrating atoms, such changes are most likely to affect the phonon modes in terms of frequency and intensity for the doped samples. Moreover, the valence state of dopant Li$^+$ is lesser than the host $Ni^{2+}$. Hence, one can expect more $O_v$ in the doped lattice than the undoped pure NiO.

The 1P(LO) mode is purely due to an O–O planar vibration in the (111) plane, while, the 1P(TO) mode arises due to O-O transverse vibrations perpendicular to the (111) plane [50-51]. As both of the 1P(TO) and 1P(LO) are correlated to the O-O vibrations and $Li^+$ doping are expected to modify the oxygen lattice, these modes are supposed to be affected. Moreover, the lesser effective mass due to $Li^+$ substitution is likely to modify the phonon frequency to higher wavenumbers, i.e., a blue shift. However, the $Li^+$-$O^{2-}$ bond may be weaker than the $Ni^{2+}$-$O^{2-}$ bond, which will be responsible for a redshift. Experimentally the 1P(TO) mode has been observed to redshift from ~381 cm$^{-1}$ (S0), to ~370 cm$^{-1}$ (S1), to ~356 cm$^{-1}$ (S3), while the 1P(LO) mode has redshifted from ~502 cm$^{-1}$ (S0), to ~482 cm$^{-1}$ (S1), to ~460 cm$^{-1}$ (S3). For the S6 sample these modes were observed ~362 cm$^{-1}$ (1P(TO)) and ~465 cm$^{-1}$ (1P(LO)) [Fig. 8 (b)]. On the other hand, the FWHM of these modes increased continuously from ~77 cm$^{-1}$ (S0) to ~86 cm$^{-1}$ (S1), to ~149 cm$^{-1}$ (S3), and ~188 cm$^{-1}$ (S6) for the 1P(TO) mode and ~132 cm$^{-1}$ (S0), to ~157 cm$^{-1}$ (S1), to ~168 cm$^{-1}$ (S3), and ~195 cm$^{-1}$ (S6) for the 1P(LO) mode [Fig. 8 (c)] [27].

The redshift of these modes indicates a stronger effect of the weakening bond strength than the effect of decreasing the effective mass of the vibrating atoms. This effect can be supported by the increasing $Ni^{2+}$-$O^{2-}$ bond length [Fig. 8 (a1)], which speaks about the decreasing bond strength of these bonds. The increasing FWHM is due to the increase of structural disorder in the lattice which is evident from the increasing Urbach energy from the UV-visible studies. Such an increase in disorder was discussed in terms of increasing $O_v$ in the lattice. From the Raman studies supported by the UV-visible studies, it is now logical to say that $O_v$ and in general O including $O_i$ play an important role in the structure of the materials, thereby modifying their vibrational and electronic properties.

A blueshift to ~465 cm$^{-1}$ for S6 can be correlated to the decrease in bond length. The blue shift for both modes observed in S6 may be a consequence of $Li^+$ being in interstitial positions, thereby reducing $O_v$ and $O_i$. Such changes introduce more complex interactions between the $Ni^{2+}$, $Li^+$, and $O^{2-}$ ions, redefining the bond length and local structure [20, 27].

A much broader and less intense 2LO mode is also observed ~ 1071.83 cm$^{-1}$ for S0, which weakened and redshifted to lower wavenumber ~ 1040.79 cm$^{-1}$ for S1 and ~ 936.21 cm$^{-1}$ for S3. The feature vanished for S6. This feature can be attributed to the Ni-O stretching vibrations corresponding to the antiferromagnetic (AFM) strength of the lattice [20, 27, 52-54]. The reduction of structural symmetry due to increasing O$_v$ proportion in the lattice near the Li$^+$ site, introduces loss of local Ni-O vibrations. The AFM ordering is greatly hampered by such disrupted lattice symmetry. On top of that, the advent of O$_i$ in S3 and more prominently in S6, adds to the degrading symmetry which is probably the reason behind the reduced AFM and thereby the vanishing of 2LO mode.

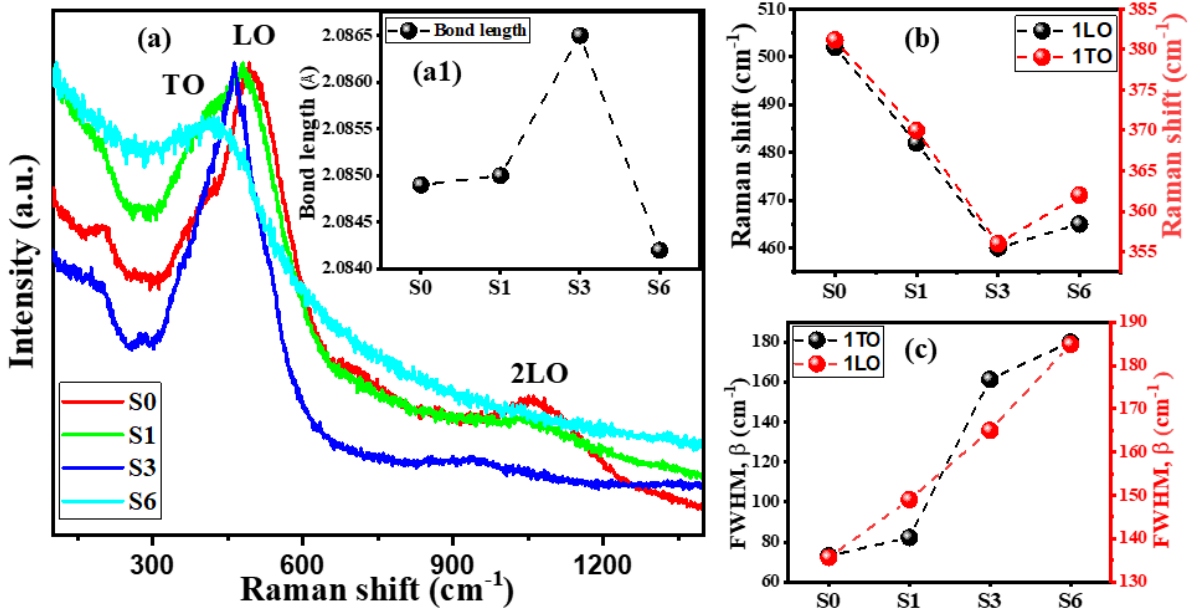

*Figure 8: (a) Raman spectra of Li-NiO samples, (b) Raman shift for 1P(LO) and 1P(TO) modes with substitution, (c) Variation of FWHM for 1P(LO) and 1P(TO) modes with substitution.*

### 3.6. Cyclic Voltammetry

The specific capacitance (C$_s$) of all samples was calculated from the CV spectra [9]:

$$C_s = (\int idt)/(\Delta V \times m \times \vartheta).$$

Here, $\int idt$ is the enclosed area under the curve, $\Delta V$ is operating potential window, $\vartheta$ is scan rate, and m is mass deposited. The energy density (E) is the energy stored per unit mass, and is defined by [9]:

$$E = \frac{1}{2}C_s \Delta V^2.$$

Oxidation and reduction peaks are related to the Faradaic reaction in the CV spectra [43, 44]:

$NiO + OH^- \rightarrow NiOOH + e^-$     (Oxidation, charging)

$NiOOH + H_2O_{ads} + e^- \rightarrow \alpha\text{-}Ni(OH)_2 + OH^-$     (Reduction, discharging)

$\alpha\text{-}Ni(OH)_2 \rightarrow NiO + H_2O$     (Dehydration in alkaline medium)

The redox peaks are observed at ~ 0.3 V (oxidation) and ~0.2V (reduction), similar to reported values [19-20]. Among the different concentrations of the active materials, the CV spectra revealed a maximum enclosed area for the 4 mg sample implying the maximum $C_s$ for this sample at 10 mV/s [Fig. 9 (a)]. Hence, a mass optimization was obtained for the 4 mg mass loading as compared to 2, 3, 5, and 6 mg mass loading. Hence, 4 mg mass loading will be used as a benchmark for further discussion in this work.

It can be observed that the CV spectra for different samples have similar profiles but with minor but different redox peak positions and enclosed areas for a specific scan rate. At a lower $\vartheta$, ions in the electrolyte have sufficient time to diffuse into deeper active sites of NiO, ensuring full redox activity. At a higher $\vartheta$, even though $\int idt$ has increased, the ratio of $\int idt/\vartheta$ decreases thereby reducing $C_s$ and hence $E$. This is because at high $\vartheta$, ions cannot penetrate deep into the electrode, thereby not allowing all active sites to participate in electrochemical reactions and only react at the outer surface [16-20].

Fig. 9 (b) shows a comparison of the CV spectra for etched NF, for S0/NF, and S3/NF [Note that the area of the CV spectra increases for etched NF as compared to the commercial and ethanol cleaned NFs, Supplementary Fig. S6]. However, this improvement is negligibly small compared to the S0/NF and S3/NF coated samples. In between the two coated samples, S3/NF sample reveals a much larger area than the S0/NF [56-60].

For reversible (diffusion-controlled) reactions, the peak current ($I_p$) follows the Randles-Sevcik equation [37-41]:

$$I_p = (2.69 \times 10^3)n^{3/2}AD^{1/2}C\vartheta^{1/2}.$$

Here, $n$ is the number of electrons transferred, $A$ is the electrode surface area, $D$ is the diffusion coefficient of the electroactive species, and $C$ is the concentration of the redox species. The value of peak potential is given according to the relation [44]:

$$E_P = E^0 + \left(\frac{0.78RT}{\alpha nF}\right)ln\vartheta.$$

Here, $E^0$ is the standard redox potential, $\alpha$ is charge transfer coefficient ($0 < \alpha < 1$), and $R, T, F$ are universal constants. Note that, $I_p$ is proportional to $\vartheta^{1/2}$, and the peak potential is proportional to $ln\vartheta$. Hence, the peak current should increase with the scan rate along with a shift in peak potential [56-60]. An increased enclosed area ($\int idt$) and peak shift are visible with increasing scan rate in Fig 9 (c) [Supplementary Fig. S7].

On the introduction of Li$^+$ in a NiO lattice, the electronic structure, conductivity, and hence the electrochemical behavior will be modified, thereby modifying the CV profile [Fig 9 (d)]. From XRD and XPS it was concluded that Li$^+$ doping increased the Ni$^{3+}$ percentage until S3. This should increase the amount Ni$^{3+}$/Ni$^{2+}$ redox couple, which should improve the electrical conductivity. The increased electrical conductivity allows faster electron transfer and therefore faster Faradaic reactions leading to higher pseudocapacitive storage. UV-vis absorption spectra revealed increased defect states inside the bandgap due to Li$^+$ incorporation, exposing more electroactive sites for redox reactions. Such increased defects state thereby facilitate the redox reactions. Hence, the increased enclosed surface area and peak potential shift can be correlated to such electronic changes leading to an improved $C_s$ from 438.86 F/g in S0, to 465.08 F/g in S1, to 528.34 F/g in S3 at 5mV/s [Fig 9 (e)]. Thereby, the value of E increases from 21.94 J/g in S0, to 23.25 J/g in S1, to 26.41 J/g in S3 [Fig 9 (f)].

For higher doping, i.e. for S6, XPS studies revealed a reduced Ni$^{3+}$/Ni$^{2+}$ ratio. Hence, $C_s$ reduced to 502.02 F/g for S6 as compared to 528.34 F/g in S3 but is higher than 465.08 F/g in S1. Similarly, E was lower (25.10 J/g) than S3 but higher than S1.

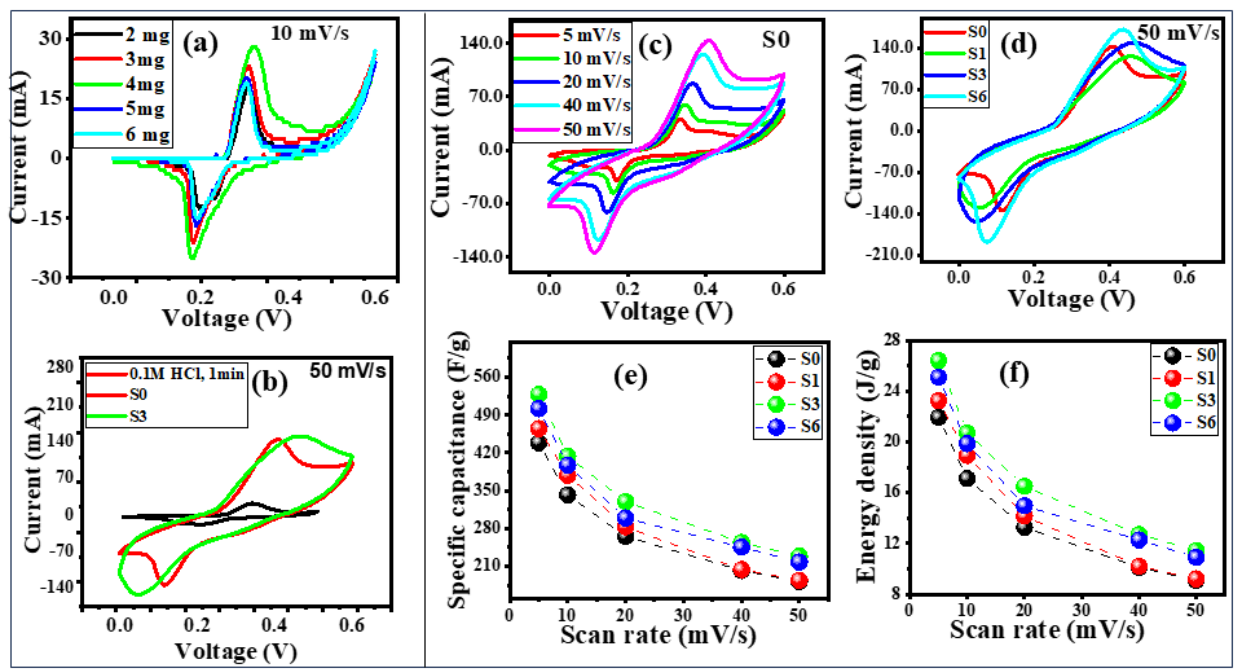

*Figure 9 (a) CV plots of mass optimization at 10 mV/s, (b) Plot comparing different CV spectra (c) CV spectra for undoped-NiO at different scan rates, (d) CV spectra for all samples at the same scan rate (50 mV/s), and (e, f) specific capacitance and energy density plot with different scan rates for all samples.*

The GCD performance at a current of 10 mA in the potential window 0–0.4 V was further studied [Fig 10 (a-b), Supplementary Fig. S8]. The $C_s$ from the GCD curve was calculated using the following relation [56-60],

$$C_s = (I\ \Delta t)/(m\ \Delta V)$$

Similar trends of $C_s$ and E were also observed from the GCD studies [Fig 10 (c-d)]. With increased $Li^+$ doping, $C_s$ and E increased for S1 (553.33 F/g and 12.26 J/g) as compared to S0 (511.54 F/g and 11.36 J/g) and further increased drastically for S3 (888.15 F/g and 19 J/g). For S6, the values of $C_s$ and E were (789.47 F/g and 17.54 J/g), i.e., lower than S3 but higher than S1. Hence, $C_s$ and E followed the same trend in both CV and GCD studies [Fig. 8 (c,d)].

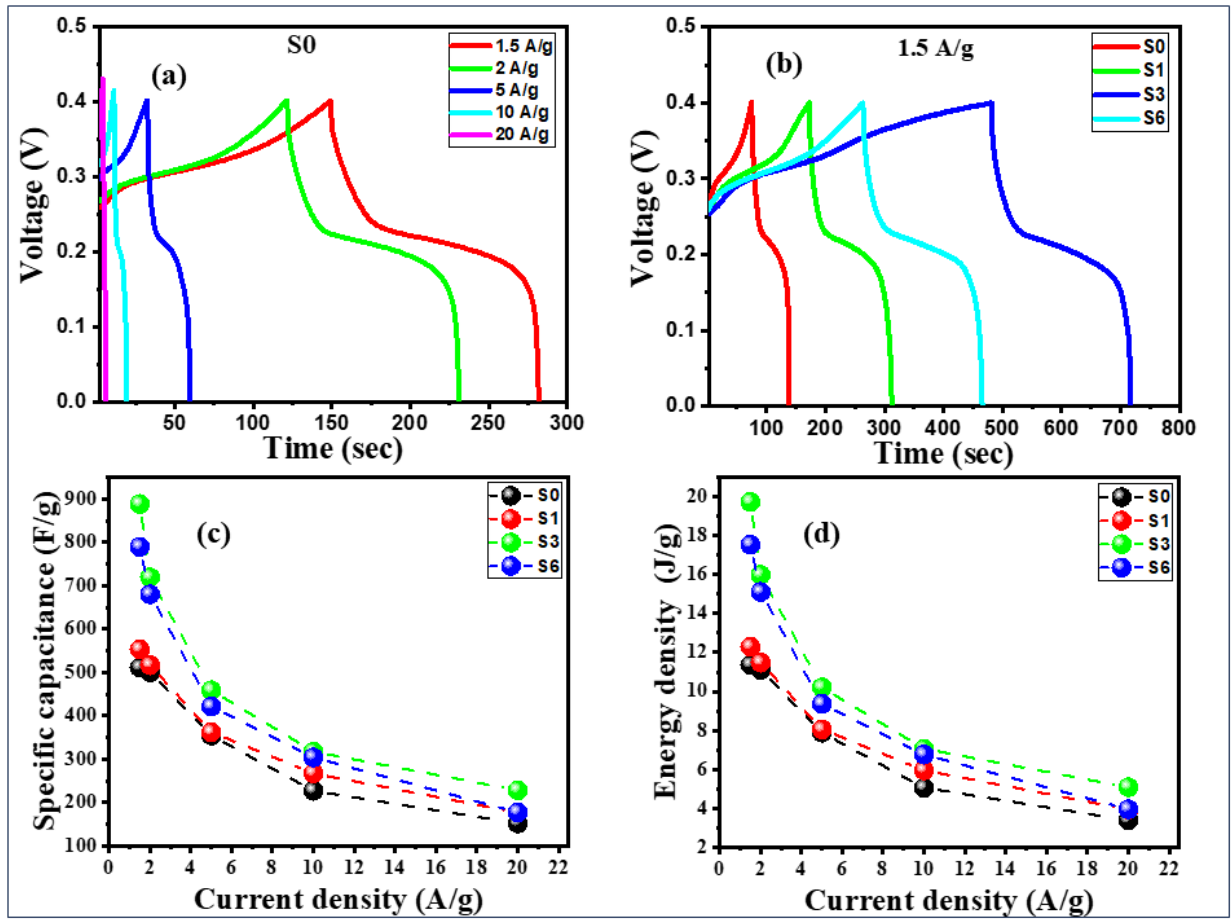

*Figure 10 (a) GCD spectra for undoped-NiO at different scan rates, (b) GCD spectra for all samples at the same scan rate (1.5 A/g), and (c, d) specific capacitance and energy density plot with different scan rates for all samples.*

Hence, this study on Li-doped NiO reveals a strong correlation between structural modifications, optical properties, electrical performance, and electrochemical properties. With $Li^+$ incorporation, the strain and disorder in the lattice increased. It was observed that the tendency of $Li^+$ to occupy the interstitial positions increases with increasing Li content. In S1, $Li^+$ mostly substitutes $Ni^{2+}$, thereby creating $Ni^{3+}$ and $O_v$. However, along with such a trend, in S3 there are 20% of $Li^+$ that are in interstitial positions. For S6, approximately 50% of added $Li^+$ is interstitial. As a result, the $Ni^{3+}/Ni^{2+}$ ratios and $O_v$ populations are proportionately altered, thereby modifying the ligand field, electronic cloud distributions, lattice vibrations, transport properties, and thereby electrochemical properties. Hence, this work is just not about a substitution of $Li^+$ in NiO lattice but a guideline on how the nature of this substitution can affect the pseudocapacitive charge storage properties of an important material like NiO.

## 4. Conclusion

$Li^+$ incorporation in a NiO lattice seems to replace $Ni^{2+}$ until 3% doping. Such a replacement simultaneously introduces a conversion of some of the host Ni ions from $Ni^{2+}$ to $Ni^{3+}$. This substitution also introduces $O_v$. Both $Ni^{3+}$ and $O_v$ creation is a consequence of charge compensation in the lattice due to a lesser valent $Li^+$ ion substitution w.r.t. $Ni^{2+}$. These changes modifying the lattice due to changes in the ligand field between $Li^+$-$O^{2-}$, $Ni^{3+}$-$O^{2-}$, $Ni^{2+}$-$O_v$, $Li^+$-$O_v$, etc. modified bonds.

Such modifications in the ligand field modified the bond strength, effective mass etc. thereby modifying the phonon frequencies and band structure. The lattice parameters decrease and lattice strain increase with such changes along with the lattice disorder. Such a modification enables the creation of enhanced charge carriers thereby improving the transport properties. This also is beneficial for improving pseudo-capacitance from 511.54 F/g in S0, to 553.33 F/g in S1, to 888.15 F/g in S3, and the energy storage properties from 11.36 J/g in S0, to 12.26 J/g in S1, to 19 J/g in S3 at 1.5 A/g. Thereby, this study shows the importance of valence state modification due to a different valence ion in the lattice on structural, electronic, transport, and energy storage properties. On the other hand, an excess of $Li^+$ ion incorporation as in 6% doped NiO results in the Li ion accepting the interstitial positions thereby reducing the $Ni^{3+}$ and $O_v$ content and having a reverse effect on the structural, electronic, transport, and energy storage properties.

**Acknowledgments**

S. Sen would like to acknowledge the Department of Science and Technology (DST), Govt of India for providing the funds (DST/TDT/AMT/2017/200). The authors also acknowledge the Department of Science and Technology (DST), Govt. of India, for providing FIST funded Raman Spectrometer (Grant Number SR/FST/PSI-225/2016) at the Department of Physics, IIT Indore. P. Maneesha would like to thank the Ministry of Education, Government of India for the Prime Minister Research Fellowship (PMRF), Govt. of India. M. Kumari would like to thank UGC for providing fellowships. A. Mekki and K. Harrabi would like to thank KFUPM for providing funds for conducting research. P. Singh acknowledges the constant moral support of Rakhi Saha, Dilip Sasmal, Prashant Joshi and Garvit Shrivastava.

# References

1. Mitali, J., S. Dhinakaran, and A. A. Mohamad. "Energy storage systems: A review." *Energy Storage and Saving* 1.3 (2022): 166-216.
2. He, Yingchun, et al. "Lattice and electronic structure variations in critical lithium doped nickel oxide thin film for superior anode electrochromism." *Electrochimica Acta* 316 (2019): 143-151.
3. Lakshmi, KC Seetha, and Balaraman Vedhanarayanan. "High-performance supercapacitors: a comprehensive review on paradigm shift of conventional energy storage devices." *Batteries* 9.4 (2023): 202.
4. Zhao, Jingyuan, and Andrew F. Burke. "Electrochemical capacitors: Materials, technologies and performance." *Energy Storage Materials* 36 (2021): 31-55.
5. Kate, Ranjit S., Suraj A. Khalate, and Ramesh J. Deokate. "Overview of nanostructured metal oxides and pure nickel oxide (NiO) electrodes for supercapacitors: A review." *Journal of Alloys and Compounds* 734 (2018): 89-111.
6. Lai, Hongwei, et al. "Mesostructured NiO/Ni composites for high-performance electrochemical energy storage." *Energy & Environmental Science* 9.6 (2016): 2053-2060.
7. Sta, I., et al. "Structural, optical and electrical properties of undoped and Li-doped NiO thin films prepared by sol–gel spin coating method." *Thin solid films* 555 (2014): 131-137.
8. Napari, Mari, et al. "Antiferromagnetism and p-type conductivity of nonstoichiometric nickel oxide thin films." *InfoMat* 2.4 (2020): 769-774.
9. Sen, Tithi, et al. "Enhanced electrochemical performance of NiO surfaces via selective Li+ doping." *Physical Chemistry Chemical Physics* 26.42 (2024): 27141-27151.
10. Egbo, Kingsley O., et al. "Efficient p-type doping of sputter-deposited NiO thin films with Li, Ag, and Cu acceptors." *Physical Review Materials* 4.10 (2020): 104603.
11. Chia-Ching, Wu, and Yang Cheng-Fu. "Investigation of the properties of nanostructured Li-doped NiO films using the modified spray pyrolysis method." *Nanoscale research letters* 8 (2013): 1-5.
12. Kate, Ranjit S., Suraj C. Bulakhe, and Ramesh J. Deokate. "Co doping effect on structural and optical properties of nickel oxide (NiO) thin films via spray pyrolysis." *Optical and Quantum Electronics* 51.10 (2019): 319.
13. Sahoo, Pooja, Akash Sharma, and R. Thangavel. "Influence of Cu incorporation on physical properties of nickel oxide thin films synthesized by sol-gel method." *AIP Conference Proceedings*. Vol. 2115. No. 1. AIP Publishing, 2019.
14. Zhang, J. Y., et al. "Electronic and transport properties of Li-doped NiO epitaxial thin films." *Journal of Materials Chemistry C* 6.9 (2018): 2275-2282.
15. Paulose, Rini, Raja Mohan, and Vandana Parihar. "Nanostructured nickel oxide and its electrochemical behaviour—A brief review." *Nano-Structures & Nano-Objects* 11 (2017): 102-111.
16. Gawali, Swati R., et al. "Asymmetric supercapacitor based on nanostructured Ce-doped NiO (Ce: NiO) as positive and reduced graphene oxide (rGO) as negative electrode." *ChemistrySelect* 1.13 (2016): 3471-3478.
17. Bharathy, G., and P. Raji. "Pseudocapacitance of Co doped NiO nanoparticles and its room temperature ferromagnetic behavior." *Physica B: Condensed Matter* 530 (2018): 75-81.
18. Yuan, Guohui, et al. "Cu-doped NiO for aqueous asymmetric electrochemical capacitors." *Ceramics International* 40.7 (2014): 9101-9105.
19. Mai, Y. J., et al. "Co-doped NiO nanoflake arrays toward superior anode materials for lithium ion batteries." *Journal of Power Sources* 196.15 (2011): 6388-6393.

20. Li, Yan, et al. "One-step synthesis of Li-doped NiO as high-performance anode material for lithium ion batteries." *Ceramics International* 42.13 (2016): 14565-14572.
21. Hu, Qiwen, et al. "Hollow Cu-doped NiO microspheres as anode materials with enhanced lithium storage performance." *RSC advances* 9.36 (2019): 20963-20967.
22. Chia-Ching, Wu, and Yang Cheng-Fu. "Investigation of the properties of nanostructured Li-doped NiO films using the modified spray pyrolysis method." *Nanoscale research letters* 8 (2013): 1-5.
23. Ranjitha, R., et al. "Rapid photocatalytic degradation of cationic organic dyes using Li-doped Ni/NiO nanocomposites and their electrochemical performance." New Journal of Chemistry 45.2 (2021): 796-809
24. Bhatt, Aarti S., et al. "Optical and electrochemical applications of Li-doped NiO nanostructures synthesized via facile microwave technique." Materials 13.13 (2020): 2961.
25. Laib, Abdellatif, et al. "Effect of Li doping on the structural, linear and nonlinear optical properties of NiO thin films." Ferroelectrics 599.1 (2022): 186-200.
26. Xu, Xianglan, et al. "Engineering $Ni^{3+}$ cations in NiO lattice at the atomic level by $Li^+$ doping: the roles of $Ni^{3+}$ and oxygen species for CO oxidation." *ACS Catalysis* 8.9 (2018): 8033-8045.
27. Mishra, Prashant Kumar, et al. "Electroluminescence, UV sensing, and pressure-induced conductance of $Li^+/Al^{3+}$ modified NiO: theoretical/experimental insights." Journal of Materials Research 38.9 (2023): 2550-2565.
28. Zhang, J. Y., et al. "Electronic and transport properties of Li-doped NiO epitaxial thin films." Journal of Materials Chemistry C 6.9 (2018): 2275-2282.
29. Jang, Wei-Luen, et al. "Electrical properties of Li-doped NiO films." *Journal of the European Ceramic Society* 30.2 (2010): 503-508.
30. Dutta, Titas, et al. "Effect of Li doping in NiO thin films on its transparent and conducting properties and its application in heteroepitaxial pn junctions." *Journal of Applied Physics* 108.8 (2010).
31. Joshi, U. S., et al. "Structure of NiO and Li-doped NiO single crystalline thin layers with atomically flat surface." *Thin Solid Films* 486.1-2 (2005): 214-217.
32. Matsubara, Kohei, et al. "Enhanced conductivity and gating effect of p-type Li-doped NiO nanowires." *Nanoscale* 6.2 (2014): 688-692.
33. Mandal, Suman, and Krishnakumar SR Menon. "Hole-states in Li doped NiO: doping dependence of Zhang-Rice spectral weight." *Physical Chemistry Chemical Physics* 26.43 (2024): 27735-27740.
34. Shannon, Robert D. "Revised effective ionic radii and systematic studies of interatomic distances in halides and chalcogenides." *Foundations of Crystallography* 32.5 (1976): 751-767.
35. Laib, Abdellatif, et al. "Effect of Li doping on the structural, linear and nonlinear optical properties of NiO thin films." Ferroelectrics 599.1 (2022): 186-200.
36. Van Elp, J., et al. "Electronic structure of Li-doped NiO." *Physical Review B* 45.4 (1992): 1612.
37. Sta, I., et al. "Hydrogen sensing by sol–gel grown NiO and NiO: Li thin films." *Journal of Alloys and Compounds* 626 (2015): 87-92.
38. Garduño, I. A., et al. "Optical and electrical properties of lithium doped nickel oxide films deposited by spray pyrolysis onto alumina substrates." *Journal of Crystal Growth* 312.22 (2010): 3276-3281.
39. Moulki, Hakim, et al. "Improved electrochromic performances of NiO based thin films by lithium addition: from single layers to devices." *Electrochimica Acta* 74 (2012): 46-52.

40. Gražulis, S., Chateigner, D., Downs, R. T., Yokochi, A. T., Quiros, M., Lutterotti, L., Manakova, E., Butkus, J., Moeck, P. & Le Bail, A. (2009). Crystallography Open Database – an open-access collection of crystal structures. *Journal of Applied Crystallography*, *42*, 726-729.
41. Bhankhar, Anupama, et al. "A Comparative Study of Experimental and Theoretical Structural Analysis of Lithium Doped Nickel Oxide Nanoparticles." *ECS Journal of Solid State Science and Technology* 12.1 (2023): 013001.
42. Xiao, Ziyi, et al. "Lithium doped nickel oxide nanocrystals with a tuned electronic structure for oxygen evolution reaction." *Chemical Communications* 57.49 (2021): 6070-6073.
43. Faid, Alaa Y., et al. "Ni/NiO nanosheets for alkaline hydrogen evolution reaction: In situ electrochemical-Raman study." *Electrochimica Acta* 361 (2020): 137040.
44. Abul-Magd, Ashraf A., H. Y. Morshidy, and A. M. Abdel-Ghany. "The role of NiO on the structural and optical properties of sodium zinc borate glasses." *Optical Materials* 109 (2020): 110301.
45. Yang, Li, and Björn Kruse. "Revised kubelka–munk theory. i. theory and application." JOSA A 21.10 (2004): 1933-1941.
46. Tauc, Jan, ed. Amorphous and liquid semiconductors. Springer Science & Business Media, 2012.
47. Elhamdi, Imen, et al. "Experimental and modeling study of ZnO: Ni nanoparticles for near-infrared light emitting diodes." *RSC advances* 12.21 (2022): 13074-13086.
48. Brik, M. G., et al. "Spin-forbidden transitions in the spectra of transition metal ions and nephelauxetic effect." *ECS Journal of Solid State Science and Technology* 5.1 (2015): R3067
49. Fromme, Bärbel. "Electronic Structure of MnO, CoO, and NiO." *dd Excitations in Transition-Metal Oxides: A Spin-Polarized Electron Energy-Loss Spectroscopy (SPEELS) Study* (2001): 5-26.
50. Trimarchi, Giancarlo, Zhi Wang, and Alex Zunger. "Polymorphous band structure model of gapping in the antiferromagnetic and paramagnetic phases of the Mott insulators MnO, FeO, CoO, and NiO." Physical Review B 97.3 (2018): 035107.
51. Al-Senani, Ghadah M., et al. "One Pot Synthesis, Surface, and Magnetic Properties of Ni–NiO@ C Nanocomposites." Crystals 13.10 (2023): 1497.
52. Trimarchi, Giancarlo, Zhi Wang, and Alex Zunger. "Polymorphous band structure model of gapping in the antiferromagnetic and paramagnetic phases of the Mott insulators MnO, FeO, CoO, and NiO." Physical Review B 97.3 (2018): 035107.
53. Luo, Xi, et al. "Lithium-doped NiO nanofibers for non-enzymatic glucose sensing." *Electrochemistry Communications* 61 (2015): 89-92.
54. Arunodaya, J., and Trilochan Sahoo. "Effect of Li doping on conductivity and band gap of nickel oxide thin film deposited by spin coating technique." *Materials Research Express* 7.1 (2019): 016405.
55. Yang, Seojin, et al. "Annealing environment dependent electrical and chemical state correlation of Li-doped NiO." *Journal of Alloys and Compounds* 815 (2020): 152343.
56. Salleh, Nor Azmira, Soorathep Kheawhom, and Ahmad Azmin Mohamad. "Characterizations of nickel mesh and nickel foam current collectors for supercapacitor application." Arabian Journal of Chemistry 13.8 (2020): 6838-6846.

57. Hu, Xiaoyan, et al. "Nickel foam and stainless steel mesh as electrocatalysts for hydrogen evolution reaction, oxygen evolution reaction and overall water splitting in alkaline media." RSC advances 9.54 (2019): 31563-31571.
58. Su, Yuefeng, et al. "Clean the Ni-rich cathode material surface with boric acid to improve its storage performance." Frontiers in Chemistry 8 (2020): 573.
59. Bakar, Nor Atikah Abu, et al. "The effect different of hydrochloric acid concentrations on the cleaning of Ni foam substrate: Structural and morphological studies." *Materials Today: Proceedings* 60 (2022): 1036-1041.
60. Xu, Qian, Jiajia Zhang, and Chunzhen Yang. "Nickel foam electrode with low catalyst loading and high performance for alkaline direct alcohol fuel cells." *Electrocatalysis and Electrocatalysts for a Cleaner Environment: Fundamentals and Applications* 4 (2022): 27.

# Supplementary file

# Correlation of the role of Li-doping in control of O-vacancies and Li-interstitial formations in NiO with electrochemical properties

Poonam Singh[a], P. Maneesha[a], Manju Kumari[a], Abdelkrim Mekki[b,c], Khalil Harrabi[b,d], Somaditya Sen[a*]

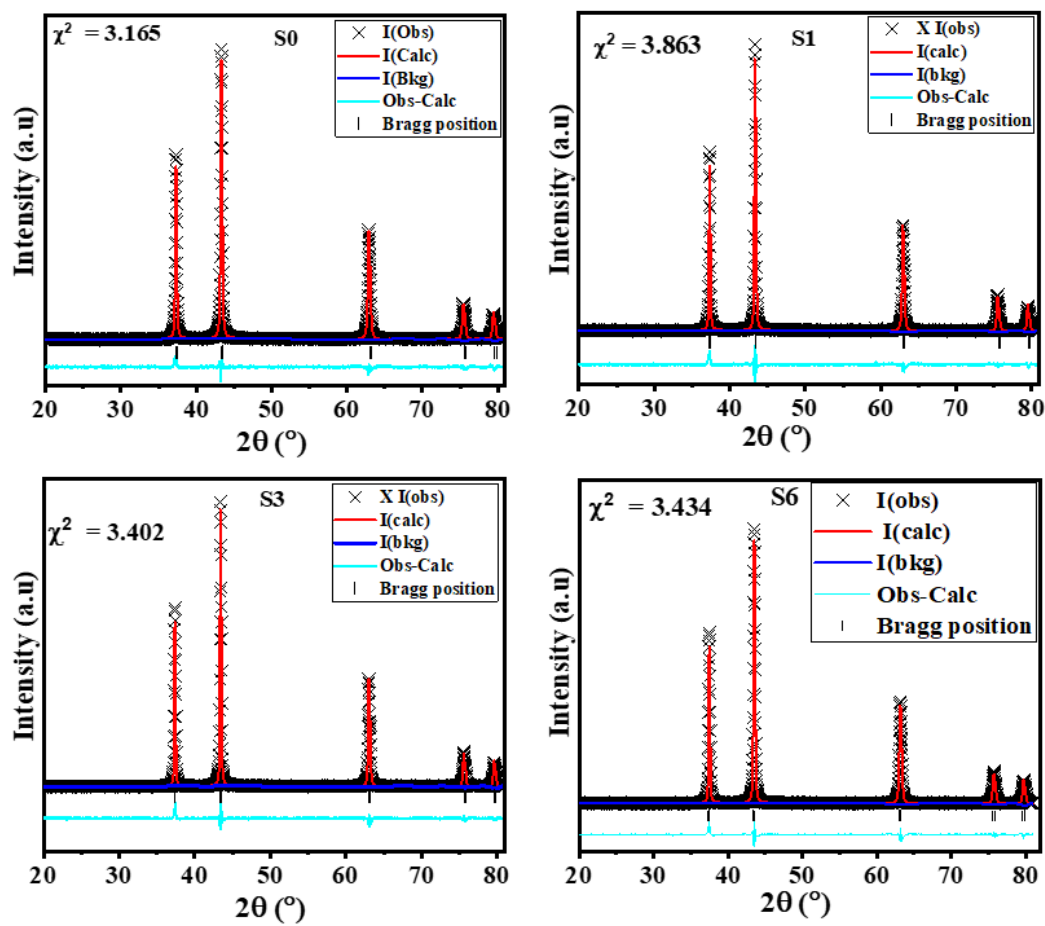

S1: Rietveld-refined patterns for all Li-doped samples.

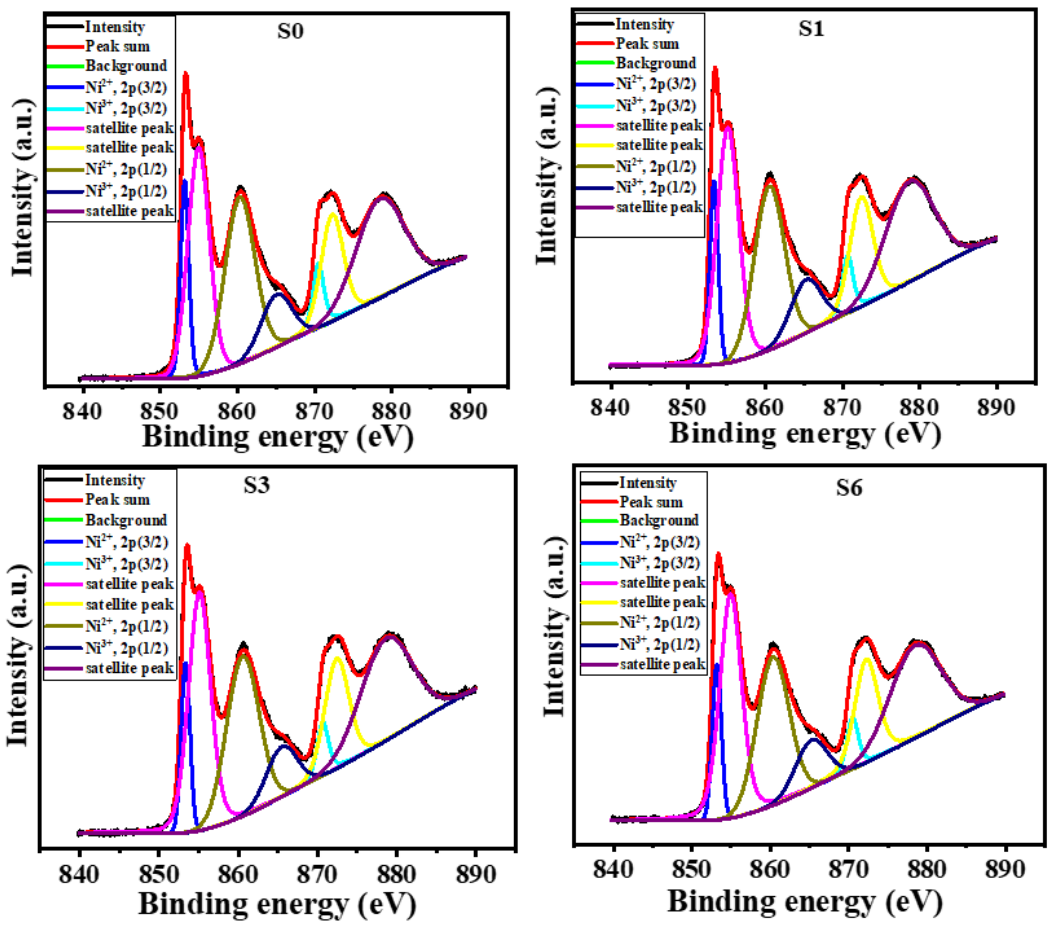

S2: Fitted XPS spectra of all Li-doped samples for Nickel component (Ni-2p).

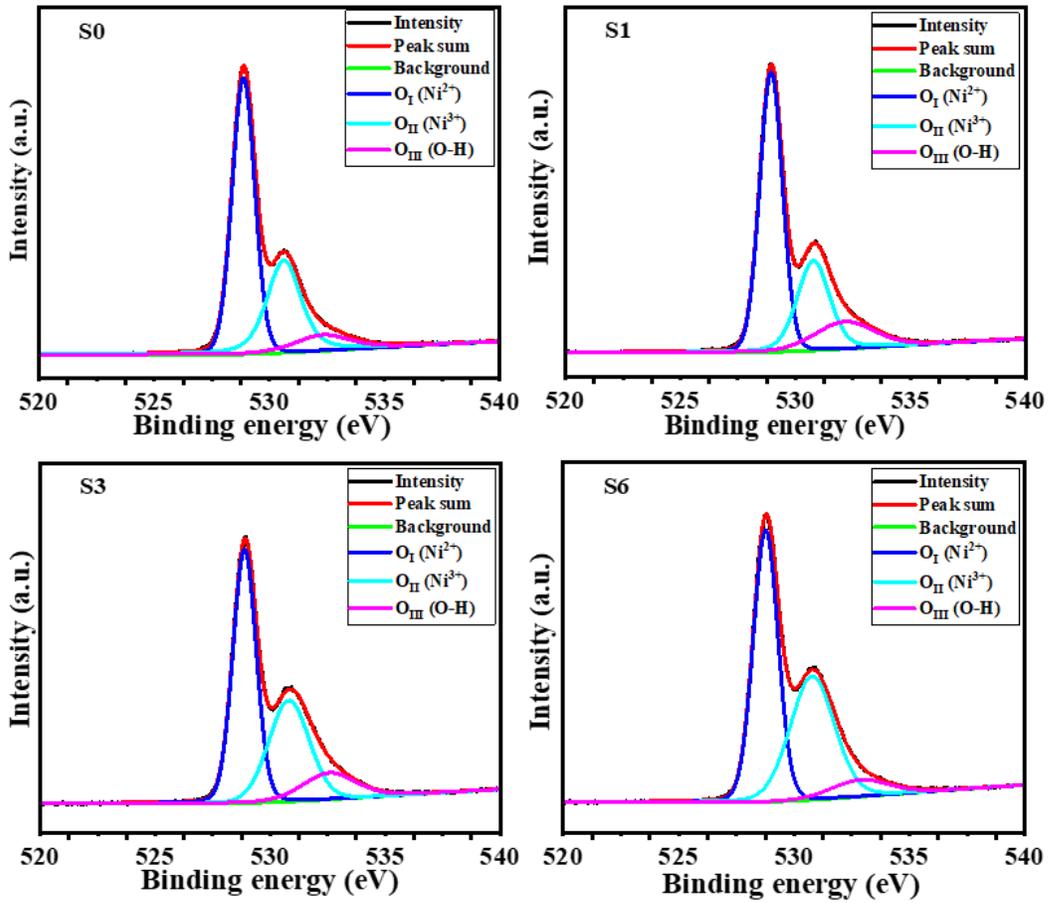

S3: Fitted XPS spectra of all Li-doped samples for Oxygen component (O-1s).

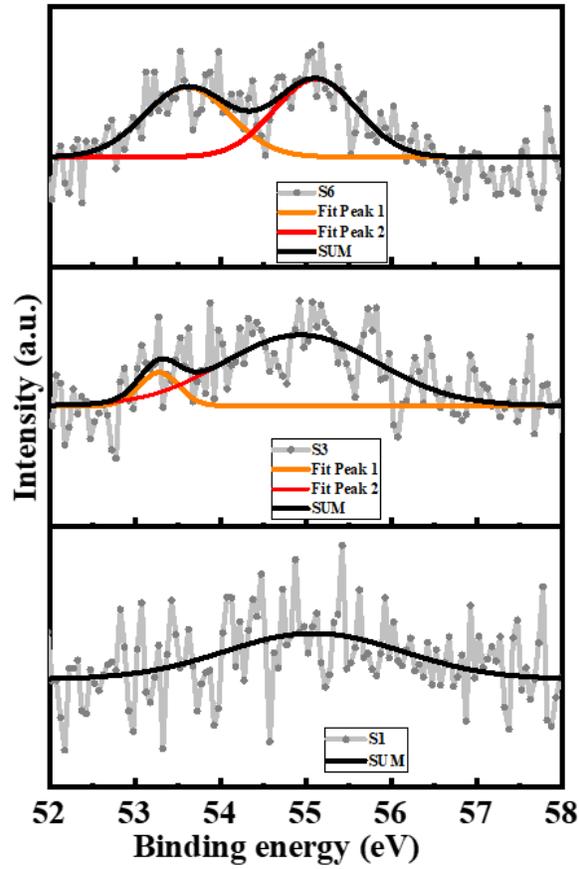

S4: Fitted XPS spectra of all Li-doped samples for lithium component (Li-1s).

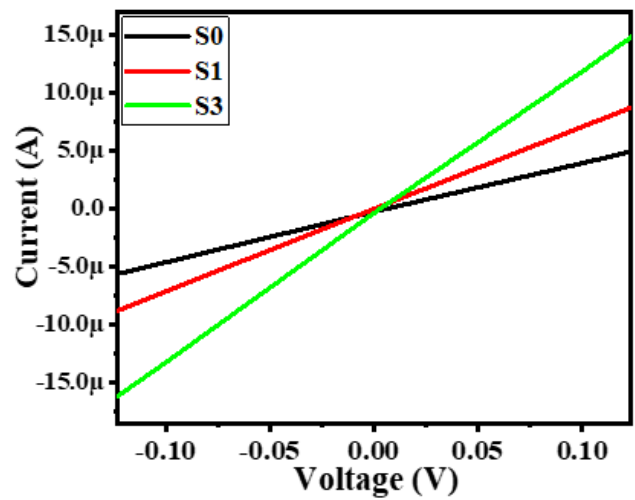

S5: IV characteristics Li-doped samples.

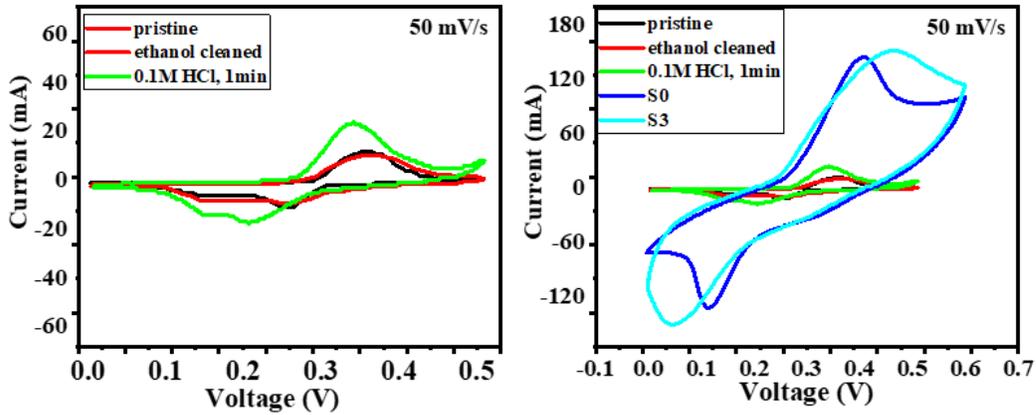

S6: (a) Plot for comparing CV spectra of pristine, ethanol, and 0.1 M HCl for 1min, (b) plot for comparing CV spectra of pristine, ethanol, and 0.1 M HCl for 1min, along with S0/NFs and S3/NFs

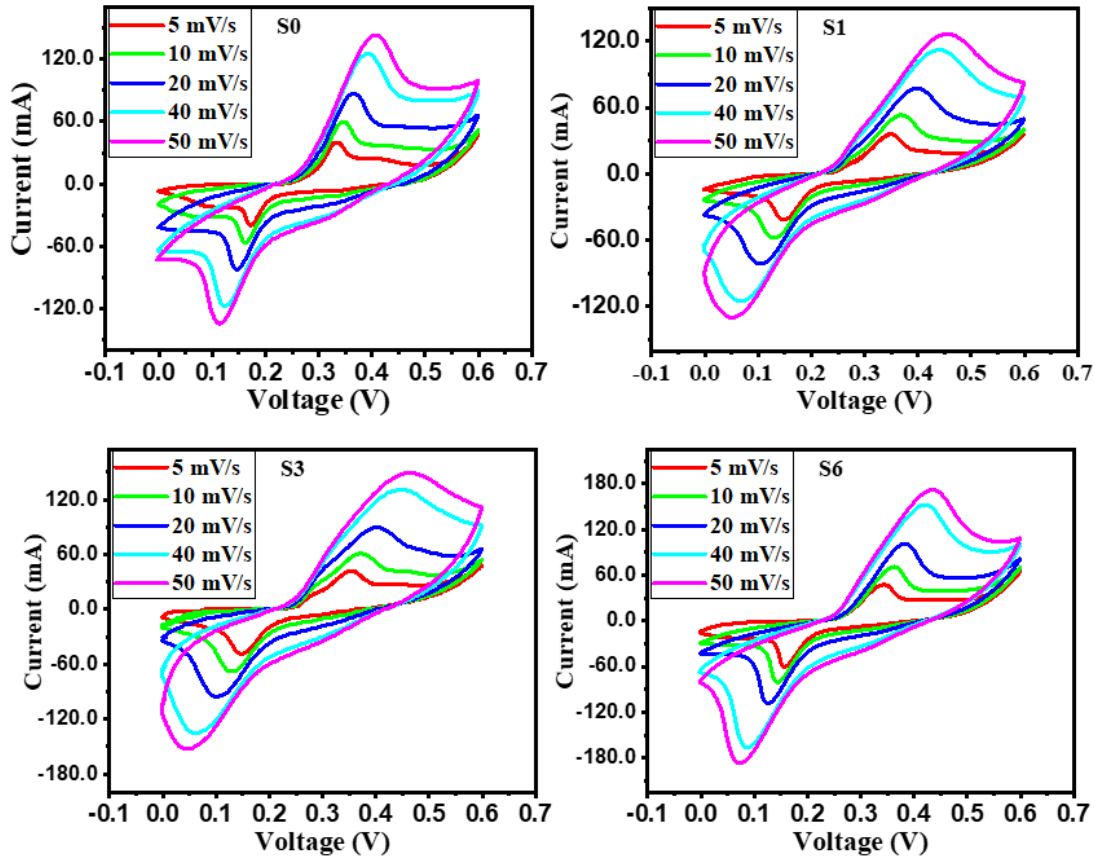

S7: (a) CV spectra of all Li-doped samples at different scan rates.

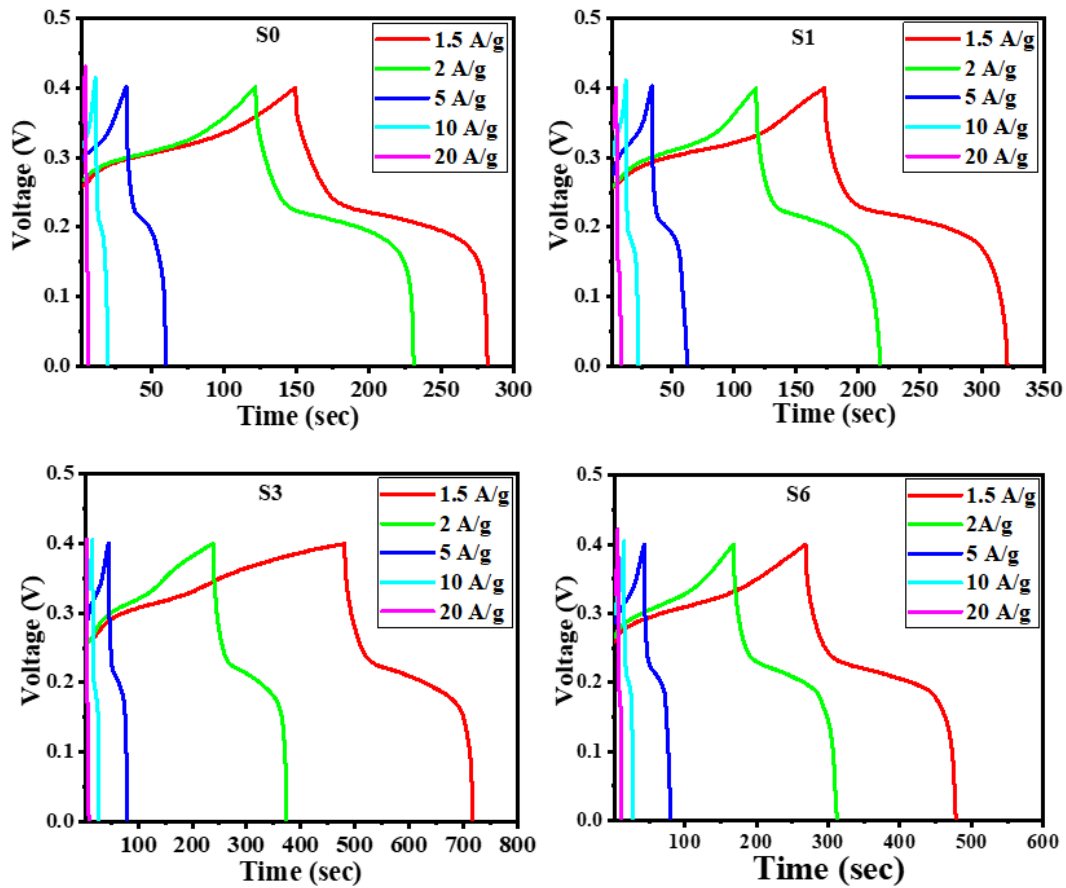

S8: (a) GCD spectra of all Li-doped samples at different scan rates.